\documentclass[USenglish]{article}	

\usepackage[utf8]{inputenc}	
\usepackage{lmodern} 
\usepackage{microtype}
\usepackage[numbers,square,sort&compress]{natbib}

\usepackage{amsthm,amsmath}
\usepackage{natbib}
\usepackage[utf8]{inputenc} 

\usepackage{lmodern} 
\usepackage{microtype}
\usepackage[numbers,square,sort&compress]{natbib}

\usepackage{amssymb}
\usepackage{graphicx,psfrag,epsf}
\usepackage{enumerate}
\usepackage{url} 
\usepackage{bm, bbm}
\usepackage{color}
\usepackage{subfig}
\usepackage{morefloats}
\usepackage{adjustbox}
\usepackage{makecell}
\usepackage{multirow}
\usepackage{tikz}
\usetikzlibrary{shapes,arrows,fit}
\usetikzlibrary{bayesnet}
\usepackage{ulem}
\usepackage{authblk}

\theoremstyle{dgthm}

\theoremstyle{dgdef}

\newcommand{\JJY}[1]{{\color{black}#1}}
\newcommand{\JJYDel}[1]{{\color{black}}}

\newcommand{\ind}{\stackrel{\rm ind}{\sim}}

\begin{document}

	\title{Bayesian modelling of response to therapy and drug-sensitivity in acute lymphoblastic leukemia}
	
	\author[1]{Andrea Cremaschi}
    \author[2]{Wenjian Yang}
    \author[1,3,4]{Maria De Iorio}
    \author[2]{William E. Evans}
    \author[2]{Jun J. Yang}
    \author[5]{Gary L. Rosner}
	\affil[1]{\protect\raggedright 
		Singapore Institute for Clinical Sciences, A*STAR, Singapore, e-mail: cremaschia@sics.a-star.edu.sg}
    \affil[2]{\protect\raggedright 
        St Jude Children's Research Hospital, Memphis, Tennessee, USA}
    \affil[3]{\protect\raggedright 
        Yong Loo Lin School of Medicine, National University of Singapore, Singapore, Singapore}
    \affil[4]{\protect\raggedright 
        Department of Statistical Science, University College London, London, UK}
    \affil[5]{\protect\raggedright 
        Johns Hopkins School of Medicine, Baltimore, Maryland, USA}

\date{}

\maketitle

\section*{Abstarct}
\noindent
Acute lymphoblastic leukemia (ALL) is a heterogeneous haematologic malignancy involving the abnormal proliferation of immature lymphocytes and accounts for most paediatric cancer cases. The management of ALL in children has seen great improvement in the last decades thanks to greater understanding of the disease leading to improved treatment strategies evidenced through clinical trials. Common therapy regimens involve a first course of chemotherapy (induction phase), followed by treatment with a combination of anti-leukemia drugs. A measure of the efficacy early in the course of therapy is the presence of minimal residual disease (MRD). MRD quantifies residual tumor cells and indicates the effectiveness of the treatment over the course of therapy. MRD positivity is defined for values of MRD greater than 0.01\%, yielding left-censored MRD observations. We propose a Bayesian model to study the relationship between patient features (leukemia subtype, baseline characteristics, and drug sensitivity profile) and MRD observed at two time points during the induction phase. Specifically, we model the observed MRD values via an auto-regressive model, accounting for left-censoring of the data and for the fact that some patients are already in remission after the first stage of induction therapy. Patient characteristics are included in the model via linear regression terms. In particular, patient-specific drug sensitivity based on \textit{ex-vivo} assays of patient samples is exploited to identify groups of subjects with similar profiles. We include this information as a covariate in the model for MRD. We adopt horseshoe priors for the regression coefficients to perform variable selection to identify important covariates. We fit the proposed approach to data from three prospective paediatric ALL clinical trials carried out at the St. Jude Children’s Research Hospital. Our results highlight that drug sensitivity profiles and leukemic subtypes play an important role in the response to induction therapy as measured by serial MRD measures.

\noindent
\textbf{keywords}: conditional modelling, left censoring, Lethal Concentration 50, minimal residual disease,  mixture models

\section{Introduction}\label{sec:Intro}

Acute lymphoblastic leukemia (ALL) accounts for around 25\% of pediatric cancers (age $\le$ 18 years) and is the most common cancer among children \cite{MarcotteEtAl2021}.
The cure rate for children with acute lymphoblastic leukemia (ALL) has improved greatly over the past few decades. Ten-year survival for childhood ALL has gone from around 11\% in the first half of the 1960s to more than 90\% by 2010 thanks to advances in understanding of the disease and clinical trials that have greatly improved outcomes \cite{PuiEvansALL, SurvivalImprovement}. 
Research in the past has revealed cytogenetic characteristics of patients' leukemic blast cells that are associated with prognosis \cite{CampanaMolecularDeterminants2008}. 
For example, children with \textit{ETV6-RUNX1}-positive ALL have a better outcome from chemotherapy than those who are negative for this fusion gene \cite{ETV6RUNX1Prognosis}. There are now many ALL subtypes defined by mutations, gene fusions as well as transcriptome with some directly informing new treatment strategies \cite{JehaEtAl2021}. 
In addition to genomic determinants of prognosis, pharmacokinetic variability, some of which is influenced by pharmacogenomics, contributes to heterogeneity in treatment response and toxicity, both systemically and at the cellular level \cite{WijayaEtAl,EstlinEtAlClinicalCellularPK2000}.

The general treatment strategy for leukemia and many other hematologic malignancies is to first reduce cancer cell burden substantially, with the objective of inducing clinical remission. This initial phase of a chemotherapeutic regimen is called the induction phase. The induction phase often includes sequential treatment with a combination of cytotoxic and/or targeted-therapies. The strategy uses intermediate indicators of treatment response after a first course of the drugs and adapts subsequent drug doses and combinations accordingly. The goal is to achieve remission by the end of the induction phase, without which the patient's prognosis is much poorer.

In ALL, the absence of minimal residual disease (MRD) has served as an early indicator of benefit from the anti-leukemia chemotherapy for decades \cite{ClinicalSignificanceOfMRD}. Methodologies for MRD measurement detect and quantify residual tumor cells beyond the sensitivity level of cytomorphology. MRD assessment can be refined by the evaluation of additional genomic markers \citep{della2019minimal}. Nowadays the clinical impact of MRD is widely accepted, and MRD is considered the most important prognostic factor in the management of ALL. The prognostic significance of MRD can vary, depending on the timing in which it is measured. Early assessment of MRD during and at the end of the induction phase is often used for clinical decision making when treating patients with ALL \cite{CampanaPuiMRDGuidedTherapy}. 

Assays for MRD have a lower limit of detection. A patient is defined as MRD negative if the patient's MRD levels fall below a detection threshold, typically less than 1/10000 cells. It is common in clinical practice to adopt a threshold of $>$ 0.01\% to define MRD positivity. This value represents a limit of detection by flow cytometry, and retrospective analyses have shown that patients whose MRD > 0.01\% after induction therapy have a greater risk of relapse and poorer prognosis \cite{campana2010minimal,BorowitzPrognosticSignificanceOfMRD2015, KruseMRD2020}. It is possible to achieve a routine sensitivity of 0.001\% by PCR in clinical samples and similar sensitivity may be achieved by flow cytometry with specific B-cell ALL subtypes \cite{TheunissenFlowMRD2017}.

The study population for this work comprises patients who participated in three clinical trials at the St. Jude Children's Research Hospital. The treatment regimens call for MRD assessments on day 15 and day 42 (end of induction) for determination of further treatment. Patients with MRD of 1\% or higher on day 15 receive more intensive therapy for the remainder of the induction phase. Further intensification is reserved for patients with 5\% of more leukemic cells on day 15. Patients with so-called standard risk ALL who have MRD of 0.01\% or higher on day 42 are reclassified as high-risk ALL. 

Additional information on the participants in the studies, collected at the beginning of the treatment, is available, such as age, gender, white blood cell count (WBC), and ALL subtype, as well as results of an \textit{ex vivo} drug sensitivity screening to characterize the patients' sensitivity profiles to anti-leukemic drugs. 

The aim of this work is to investigate the relationship between patients' leukemia subtype, leukemic cell drug sensitivity, and clinical benefit as measured by the early marker of treatment effect (MRD). To this end, we specify a joint model in a Bayesian framework for the MRD outcomes collected at days 15 and 42 able to account for the censoring and time dependence between the MRD measurements.
Our statistical model combines multiple complexities and allows inference from the full Bayesian model. The complexities include the following. The model allows for the presence of leukemic cells below the assay's lower limit of detection by treating MRD assessments as left censored if recorded as $<$0.01\%. The model also considers MRD assessments on days 15 and 42 jointly via an autoregressive model. For day 42 MRD, inference is conditional on the presence of MRD on day 15, since no patient who is MRD negative on day 15 becomes MRD positive on day 42. This conditional autoregressive model allows for clearer interpretation of covariate effects. The model also includes Bayesian clustering as part of posterior fitting. The clustering of patients is based on \textit{ex vivo} measurements of the sensitivity of each patient's own leukemic cells to the anticancer drugs the patient receives during induction. The investigation also includes the genomic ALL subtype of each patient. A final component of the model is the use of the horseshoe prior \cite{CarvalhoEtAlHorseshoe2010} for selecting important covariates from the large number of features examined. This class of prior distributions for the regression coefficients provides a way to identify important covariates, thanks to the strong shrinkage effect of its heavy tails. In particular, the density function of this distribution presents a singularity at zero,  leading to a more aggressive shrinkage of small coefficients towards zero  than other standard prior distributions, and leaving important larger coefficients unaffected. A performance comparison between the horseshoe distribution and  the spike-and-slab approach of \cite{george1993variable} is offered in \cite{CarvalhoEtAlHorseshoe2010}, showing consistency of the two posterior variable selection results. For further discussion on prior elicitation for the horseshoe distribution, see \cite{piironen2017sparsity}.

The manuscript is organized as follows. Section~\ref{sec:Data} describes the  ALL data, while  Section~\ref{sec:Model} introduces the modelling strategy. In Section~\ref{sec:Application} we present the analysis results. Section~\ref{sec:Conclusion} concludes the paper with a discussion.

\section{ALL Dataset}\label{sec:Data}

The motivating application consists of data for patients treated in three prospective pediatric ALL clinical trials at the St. Jude Children's Research Hospital. These studies were part of the long-standing Total Therapy clinical trials program at the institution, initiated in 1962, comprising clinical trials that continue to build sequentially on each other \cite{TotalTherapyStudies}. These studies have shown the benefit of full-dose chemotherapy and treatment directed at the central nervous system for treating pediatric ALL, with greatly improved outcomes over the years, currently achieving event-free survival that exceeds 90\% \cite{PuiEvansALL}. 

Our data set includes MRD measurements for childhood ALL patients in Total Therapy studies XV ($n = 192$), XVI ($n = 428$), and XVII ($n = 168$). The treatment regimens for the studies include many of the same drugs during the six-week induction phase. There are several anti-leukemic agents that are common to the three studies; prednisone, vincristine, daunorubicin, PEG-asparaginase, methotrexate, cyclophosphamide, cytarabine, and mercaptopurine. MRD measurements are made on day 15 and at the end of the remission induction phase of treatment (day 42). 

\begin{figure}[ht]
	\centering
	\includegraphics[scale=0.75]{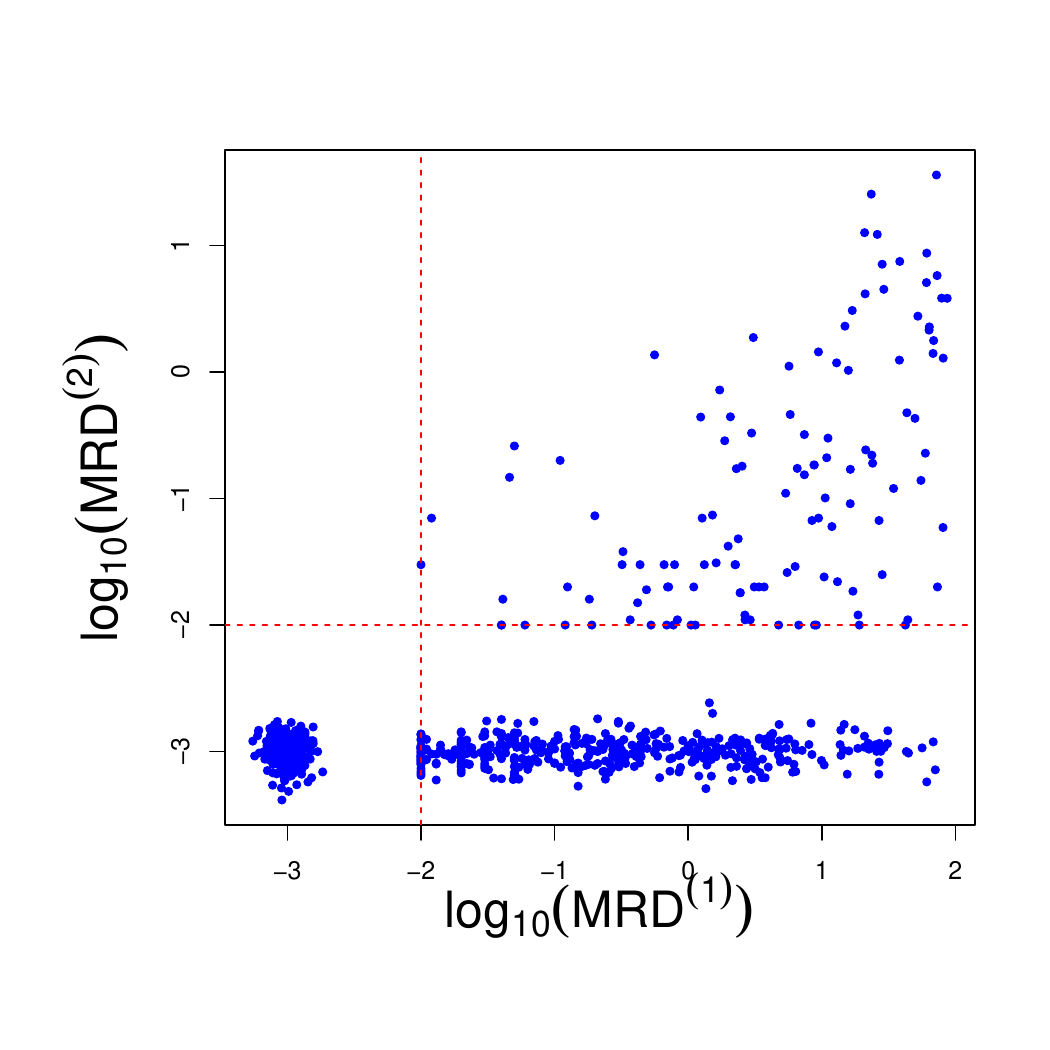}
	\caption{Observed MRD values on $\log_{10}$ scale at days 15 and 42. Values smaller than 0.01 are represented as noisy observations centred at $\log_{10} 0.001 =-3$.}
	\label{fig:MRD_obs_jitter}
\end{figure}

The detection limit for MRD is 0.01\% (1 leukemia cell among 10,000 normal cells in the bone marrow), below which the exact values are not considered accurately measurable. Figure \ref{fig:MRD_obs_jitter} displays the observed MRD values at day 15 and 42 (on the $\log_{10}$ scale). The values of MRD smaller than 0.01 are therefore censored and are jittered around 0.001 in the figure to improve presentation. The detection limit for MRD poses  a modelling challenge, which we address in this work.
The day 15 MRD assessment determines subsequent treatment during the remainder of the induction phase. Additionally, patients whose leukemic cells exhibit genomic variants for which targeted agents are available receive these chemotherapeutic agents (e.g. a tyrosine kinase inhibitor for patients whose ALL cells harbor a chromosomal translocation creating a BCR/ABL fusion). In our analysis, we focus on the drugs that all patients receive.

Additional information recorded at entry include age, gender, white blood cell count (WBC), ALL subtype, and treatment protocol number. Characteristics of the patients that are available across the three studies are summarised in Table \ref{tab:Table1}. The median age of the children is between 4.9 years and 5.6 years, typical for childhood ALL. There are slightly more male than female patients. Baseline white blood cell counts are somewhat lower in Total XVI than in the other two studies.

\begin{figure}[ht]
	\centering
	\includegraphics[scale=0.75]{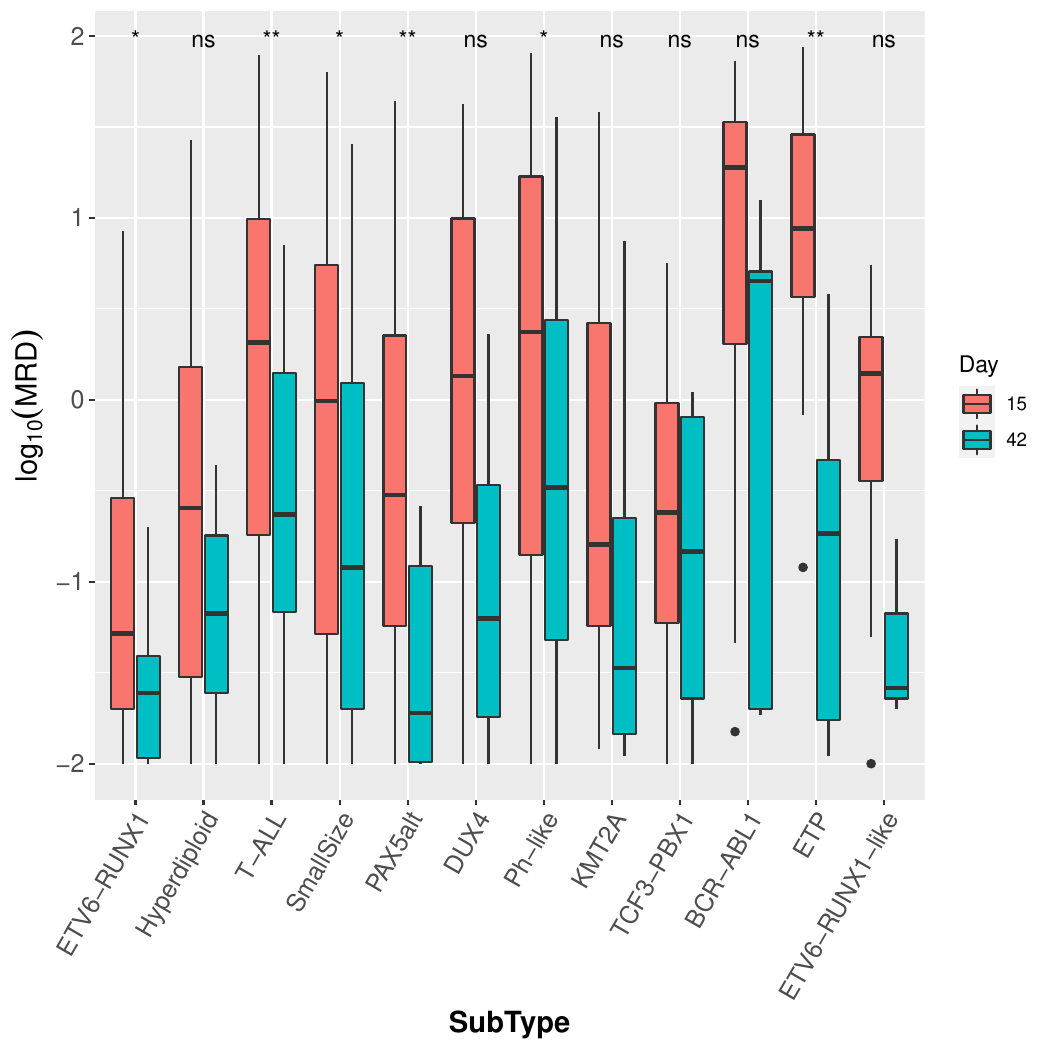}
	\caption{Observed MRD values on $\log_{10}$ scale at days 15 and 42, for each subtype category.}
	\label{fig:MRD_by_subtype}
\end{figure}

\begin{figure}[ht]
	\centering
	\includegraphics[scale=0.75]{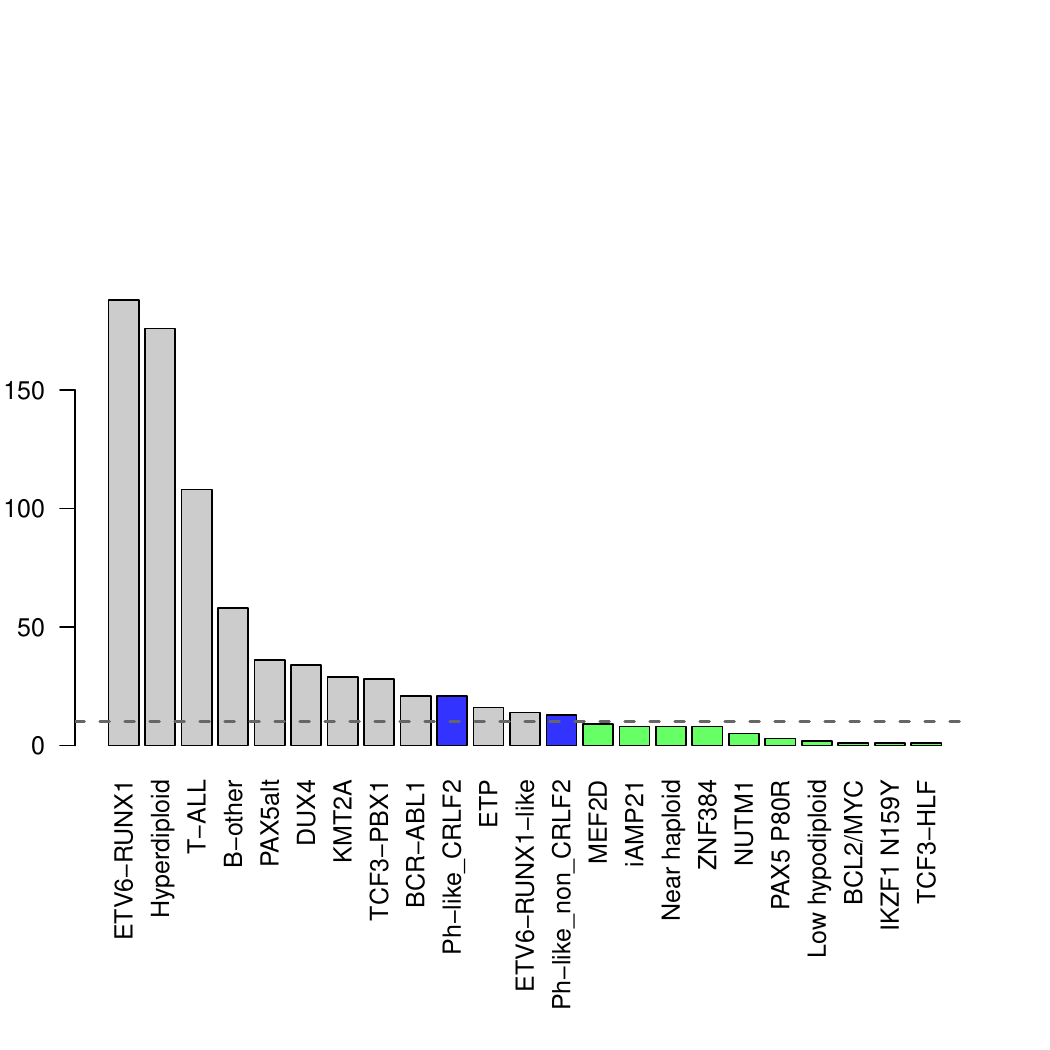}
	\caption{Distribution of the available subtypes. Grey bars correspond to subtypes modelled individually, blue to the Ph-like subtypes and green to the subtypes with less than 10 observations. Within the last two classes, the subtypes are merged before performing the analysis.}
	\label{fig:Subtypes}
\end{figure}

The genomic subtype categories are shown in Figure \ref{fig:Subtypes}. In the data set, 23 subtypes are observed, with the most common subtypes being hyperploid and \textit{ETV6-RUNX1}, typical for childhood ALL. Section \ref{sec:Intro} includes information about the fusion product \textit{ETV6-RUNX1}, which confers a favorable prognosis. Hyperdipoid ALL is also a common subtype that is also considered to have a more favorable prognosis \cite{PaulssonHyperdiploidALL}. 
We merged several subtypes that included fewer than 10 subjects each to improve degrees of freedom for this analysis. For example, we merged the Ph-like CRLF2 and the Ph-like non-CRLF2 subtypes \cite{TasianEtAlPhLike2017}. The analysis included 12 distinct subtypes. The detected MRD levels vary for different genomic categories, as shown in Figure \ref{fig:MRD_by_subtype}, displaying the measured MRD values on a $\log_{10}$-scale at each time point and across subtypes.

\begin{figure}[ht]
	\centering
	\includegraphics[width=0.75\textwidth]{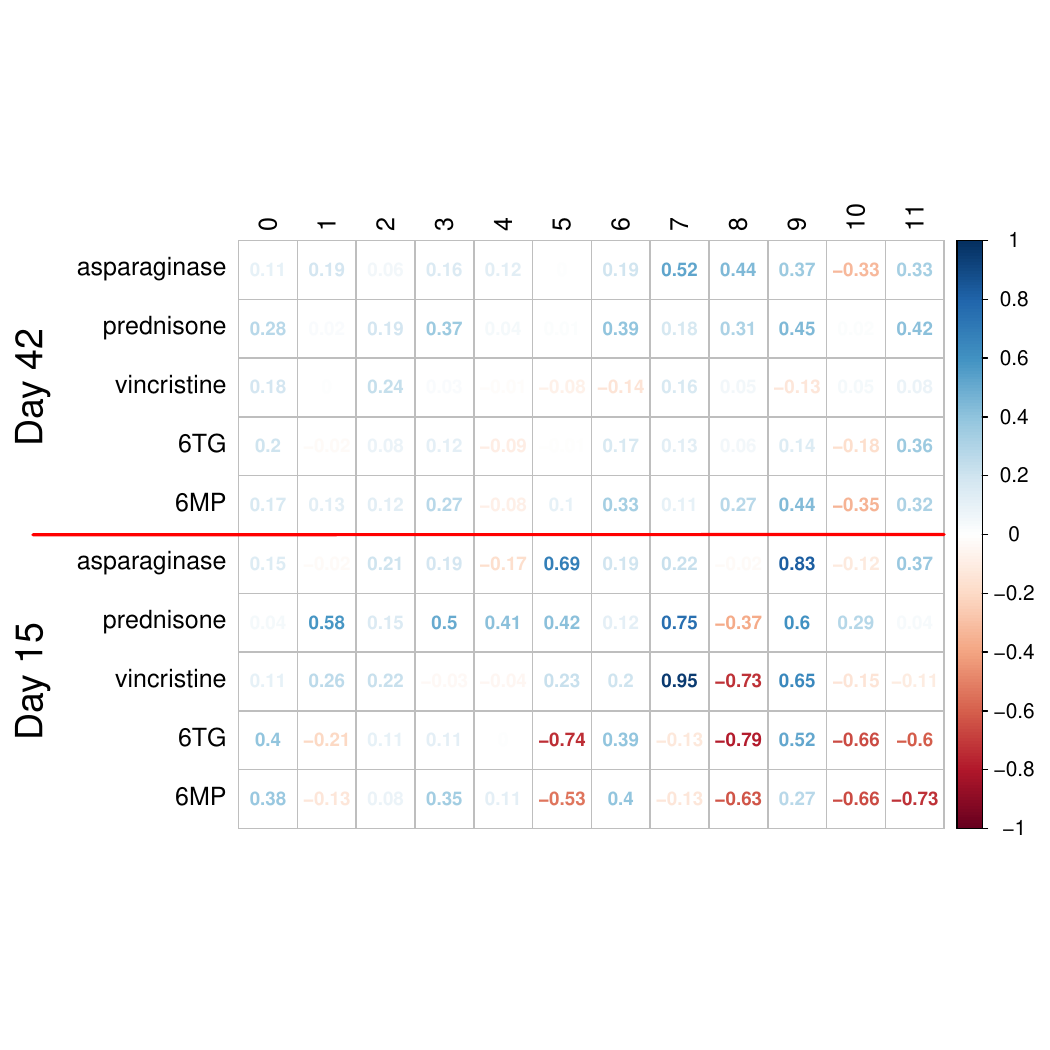}
	\caption{Pearson correlations between $\log_{10}$-MRD values and estimated LC$_{50}$s for the five compounds used in the analysis. The correlations are computed for 100 imputed $\log_{10}$ LC$_{50}$ data sets, and then averaged at each time point and for each subtype.}
	\label{fig:Pearson_Corrs}
\end{figure}

The data set includes estimates of patient-specific cancer cell sensitivities to various drugs in the combination chemotherapeutic regimen. The estimates arise from \textit{ex vivo} dilution assays of patient samples. In essence, the patient's leukemic cells are plated with each drug via dilution assays to determine sensitivity. For each drug concentration, a count of the number of surviving cells provides an estimate of the lethality of that concentration to the patient's leukemia cells. From the roughly six concentrations per drug, an estimated LC$_{50}$, the concentration that led to 50\% of the cells dying, is used as the measure of the sensitivity of each patient's disease to the specific drug. We included five anti-leukemia drugs in these analyses, selected after excluding the compounds for which the missing rate is more than 40\%. The compounds used for this analysis are asparaginase, prednisone, vincristine, 6TG and 6MP. The entire ALL pharmacotype dataset was previously published \cite{StJudePaper}. To highlight the relationship between the observed MRD values and the estimated LC$_{50}$s, in Figure \ref{fig:Pearson_Corrs}  we present the Pearson correlations between these two variables for each of the five compounds and for each subtype category, averaged over 100 imputed $\log_{10}$ LC$_{50}$ data sets (see Section~\ref{sec:Model}).

Estimates of LC$_{50}$s for patients are not available for some medications if the number of available cells was not sufficient to test all medications. Therefore, we applied multiple imputation for the missing observations. We impute missing data via the R package \texttt{mice} \cite{van2011mice}.

\begin{table}[!ht]
	\caption[]{Patient characteristics across the three studies.}\label{tab:Table1}
	\begin{tabular}{lcccc}\hline
		 & Total XV & Total XVI & Total XVII & Overall \\
		 & (N = 192) & (N = 428) & (N = 168) & (N = 788) \\\hline
		age &  &  &  & \\
        \quad Mean (SD) & 6.45 (4.36) & 7.13 (4.82) & 6.89 (4.60) & 6.92 (4.67) \\
        \quad Median [Min, Max] & 4.89 [1.02, 18.7] & 5.63 [0.120, 18.9] & 5.58 [1.00, 18.5] & 5.42 [0.120, 18.9] \\
        gender &  &  &  & \\
        \quad Female & 93 (48.4\%) & 182 (42.5\%) & 81 (48.2\%) & 356 (45.2\%) \\
        \quad Male & 99 (51.6\%) & 246 (57.5\%) & 87 (51.8\%) & 432 (54.8\%) \\
        WBC at diagnosis &  &  &  & \\
        \quad Mean (SD) & 56.1 (105) & 46.3 (89.5) & 59.4 (105) & 51.5 (96.8) \\
        \quad Median [Min, Max] & 17.4 [1.20, 1010] & 14.2 [0.700, 638] & 20.2 [1.30, 730] & 16.6 [0.700, 1014] \\
        MRD (day 15)  &  &  &  & \\
        \quad Mean (SD) & 3.50 (10.7) & 4.67 (13.3) & 3.62 (11.7) & 4.16 (12.4) \\
        \quad Median [Min, Max] & 0.0295 [0, 72.1] & 0.116 [0, 86.8] & 0.0320 [0, 80.7] & 0.0500 [0, 86.8] \\
        MRD (day 42) &  &  &  & \\
        \quad Mean (SD) & 0.460 (2.90) & 0.131 (0.846) & 0.167 (1.96) & 0.219 (1.81) \\
        \quad Median [Min, Max] & 0 [0, 36.0] & 0 [0, 12.2] & 0 [0, 25.4] & 0 [0, 36.0] \\
        MRD cat. (day 15) &  &  &  & \\
        \quad $<0.01$ & 69 (35.9\%) & 129 (30.1\%) & 65 (38.7\%) & 263 (33.4\%) \\
        \quad $0.01-1$ & 83 (43.2\%) & 161 (37.6\%) & 58 (34.5\%) & 302 (38.3\%) \\
        \quad $1-5$ & 13 (6.8\%) & 76 (17.8\%) & 25 (14.9\%) & 114 (14.5\%) \\
        \quad $>=5$ & 27 (14.1\%) & 62 (14.5\%) & 20 (11.9\%) & 109 (13.8\%) \\
        MRD cat. (day 42) &  &  &  & \\
        \quad $<0.01$ & 152 (79.2\%) & 365 (85.3\%) & 149 (88.7\%) & 666 (84.5\%) \\
        \quad $0.01-1$ & 27 (14.1\%) & 49 (11.4\%) & 18 (10.7\%) & 94 (11.9\%) \\
        \quad $>=1$ & 13 (6.8\%) & 14 (3.3\%) & 1 (0.6\%) & 28 (3.6\%) \\\hline
	\end{tabular}
\end{table}

The main goals of the statistical model are: (i) to characterize the MRD responses at both time points accounting for their time dependence and the censoring; (ii) to investigate the effect of baseline and in-trial covariate information on the MRD levels; (iii) to identify subgroups of patients presenting similar sensitivity to drug exposures and (iv) to investigate the relationship between the clustering structure and drug-subtype interactions. 

\clearpage
\section{Model specification}\label{sec:Model}

 Data are available for $N = 788$ children, for whom both baseline and in-trial information are available. Covariates included in the analysis are age, gender, WBC at diagnosis, ALL subtype and therapy protocol, as well as drug sensitivity profiles obtained from \textit{ex vivo} dilution assays. The regression model included eleven indicator variables for the twelve ALL subtypes. For each patient, at most one of these eleven parameters will equal 1. The baseline category is \textit{ETV6-RUNX1}, the most prevalent subtype. If a patient is positive for \textit{ETV6-RUNX1}, all eleven subtype variables will be 0. 
 
 The ALL data set presents two main  statistical challenges: (i) censoring of the MRD levels due to the assay's lower limit of detection (i.e., $<$ 0.01\%); (ii) estimation of each patient's drug sensitivity profile. 
As stated above, MRD values above the threshold 0.01\% are observed. We chose to treat the lack of MRD as a censored observation instead, allowing for the fact that there may still be residual disease but at an undetectable level. This consideration is particularly important when introducing temporal dependence. In the likelihood, censoring can be accounted for using standard strategies from survival analysis and censored observations are imputed as part of the Markov chain Monte Carlo (MCMC) algorithm. 

We account for the repeated measures aspect of the MRD assessments by including an autoregressive term (of order 1) to model dependence between the MRD at day 15 and day 42, only for those observations for which MRD at day 15 is above the detection threshold. Otherwise, the two time points are treated as conditionally independent. We opt for this strategy because in the dataset if MRD at day 15 is censored (i.e., $<$ 0.01\%), then MRD at day 42 also falls below the detection threshold (see Figure~\ref{fig:MRD_obs_jitter}). Introducing dependence among the observations censored at both times would bias the estimate of the temporal effect as well as of the covariate effects, since the statistical model includes covariate information via a linear regression term.
 

Another important feature of our modelling strategy is the inclusion  of a patient-specific measure of the patient's leukemic cells' sensitivity to individual drugs in our model of MRD. Drug sensitivity profiles for each patient are estimated from  \textit{ex vivo} drug sensitivity assays. The model includes a finite mixture model for the profiles to allow the data to perhaps determine categories that relate to overall sensitivity or resistance to one or more of the anticancer drugs. A major advantage  of the Bayesian framework is that it allows for joint estimation of the MRD model and the drug sensitivity profiles, enabling a probabilistically sound quantification and propagation of uncertainty.

 
 We describe in detail the statistical strategy below. 

\paragraph*{Modelling censored MRD observations}
Let $MRD^{(1)}, MRD^{(2)}$ denote the minimal residual disease measured at day 15 and 42 from entry into the study, respectively. Let $\bm Z_i = (Z^{(1)}_i, Z^{(2)}_i) := \log_{10}\left(\text{MRD}^{(1)}, MRD^{(2)}\right)$.  
Let $f\left(z^{(t)}_i \mid \bm X_i, \bm c_i, \bm \theta^{(t)}\right)$ be the conditional density function of the $i$-th observation at time $t$ (= 1, 2), given the covariates $\bm X_i$, the drug sensitivity profiles $\bm c_i$ and the vector of parameters $\bm \theta^{(t)}$. Similarly, let $F^{(t)}\left(z \mid \bm X_i, \bm c_i, \bm \theta^{(t)}\right)$ be the conditional distribution function corresponding to the density $f$ and denoting the probability of the response falling below a value $z$. Recall that the MRD values below 0.01 are censored at the lower detection level. Let $z_\text{low}$ be this threshold value.  The probability of observing a value lower than the detection threshold is, therefore:
\begin{equation}\label{eq:Prob_Cens}
	F^{(t)}(z_\text{low}) = \mathbb{P}(Z^{(t)} \leq z_\text{low}\mid \bm X_i, \bm c_i, \bm \theta^{(t)}), \quad t = 1, 2, \quad z_\text{low} = \log_{10}(0.01) = -2
\end{equation}

 We assume that $MRD^{(1)}$ and $MRD^{(2)}$ are conditionally independent given covariate information, the drug sensitivity profiles, and the parameters in the model. Let $\delta^{(t)}_i$ be the left-censoring indicator at time  $t = 1, 2$ for subject $i = 1, \dots, N$, where $\delta^{(t)}_i = 0$ if the corresponding MRD value is below the detection threshold (i.e., censored) and $\delta^{(t)}_i = 1$ if it is observed. The likelihood of the model is then:
\begin{gather}\label{eq:Likelihood}
	L(\bm z \mid \bm \theta) = L(\bm z^{(1)} \mid \bm \theta^{(1)})L(\bm z^{(2)} \mid \bm \theta^{(2)}) = \nonumber \\ 
	\prod_{t = 1}^2 \left(\prod_{i = 1}^N f\left(z^{(t)}_i \mid \bm X_i, \bm c_i, \bm \theta^{(t)}\right)^{\delta^{(t)}_i}F^{(t)}\left( z_\text{low} \mid \bm X_i, \bm c_i, \bm \theta^{(t)}\right)^{1 - \delta^{(t)}_i}\right)
\end{gather}
We specify a Normal distribution as $f\left(Z^{(t)}_i \mid \bm X_i, \bm c_i, \bm \theta^{(t)}\right) $ to model the $\log_{10}$ MRD observations and assume   
	\begin{align}\label{eq:MRD_model}
	& Z^{(t)}_i \mid \mu^{(t)}_i, \sigma^{2(t)} \sim \text{N}\left(\mu^{(t)}_i, \sigma^{2(t)}\right), \quad t = 1, 2 \nonumber \\
	& \mu^{(1)}_i = \beta^{(1)}_0 + \bm \beta^{(1)} \bm X_i + \bm \gamma^{(1)} \bm c_i \\
	& \mu^{(2)}_i = \beta^{(2)}_0 + \delta^{(1)}_i \left(
	\rho_0 + \rho Z^{(1)}_i + \bm \beta^{(2)} \bm X_i + \bm \gamma^{(2)} \bm c_i \right) \nonumber \\
	&\bm \beta^{(t)} \sim \text{HS}, \quad t = 1, 2 \nonumber \\
	& \bm \gamma^{(t)} \sim \text{HS}, \quad t = 1, 2 \nonumber \\
	&\rho_0 \sim \text{N}(0,1) \nonumber \\
	&\rho \sim \text{N}(0,1) \nonumber \\
	&\sigma^{2(t)} \sim \text{Inv-Gamma}(3, 2), \quad t = 1, 2 \nonumber
	\end{align}
where $\text{N}(\mu, \sigma^2)$ is the normal distribution with mean $\mu$ and variance $\sigma^2$, $\text{HS}$ represents the Horseshoe prior \cite{CarvalhoEtAlHorseshoe2010} for the regression parameters following the specification in \citep{piironen2017sparsity} and $\text{Inv-Gamma}(a, b)$ is the inverse Gamma distribution with  mean $b/(a-1)$ and variance $b^2/((a-1)^2(a-2))$. The HS prior belongs to the family of continuous shrinkage prior distributions, characterized by a singularity at zero, but whose fat tails still allow for large values of the coefficients.  
For each subject $i$, the covariate vector $\bm X_i$ contains information on age, gender, WBC (on $\log_{10}$ scale), ALL subtype and therapy protocol. The variable $\bm c_i$ contains information on the drug sensitivity profile (i.e., cluster membership) of the $i$-th subject and is based on the results of the \textit{ex vivo} assays. Estimation of $\bm c_i$ is a key  component of the proposed modelling strategy and will be discussed later in more details.

Note that the specification of the mean terms $\mu^{(1)}_i$ and $\mu^{(2)}_i$ is different. For day 15, $\mu^{(1)}_i$ contains standard regression terms on patients covariates, including ALL subtype and the clusters from the sensitivity profiles. At time $t = 2$, however, the mean term contains an auto-regressive term ($\rho$) along with regression coefficients, and estimation only occurs if the MRD level at time $t=1$ is not censored (i.e., $\delta^{(1)}_i = 1$). By doing so, the treatment effect at day 15 is taken into account only if the patient has detectable MRD on day 15 (right quadrant of Figure \ref{fig:MRD_obs_jitter}). Indeed, in our data set if MRD is not detected at day 15, it is also not detected at day 42. The regression results would be biased by the imputation of the censored observations (which do not contain much information on covariate effects at day 42).

\paragraph*{Drug sensitivity profiles} 

In this section we describe how the results from the \textit{ex vivo} drug sensitivity assays (see Section \ref{sec:Data}) are used to estimate drug sensitivity profiles. Here drug sensitivity is estimated by the  concentration of a drug that kills half of the tested cells in culture (LC$_{50}$).
Let $\bm Y_i$ be the patient $i$'s vector of $\log_{10}$(LC)$_{50}$ values for the panel of $d = 5$ anti-leukemic drugs. Our goal is to cluster individuals based on their  cancer cells' sensitivity to these five drugs. We perform an initial analysis of the drug sensitivity profiles to determine how many clusters are supported by the data.  Because the missing rate is high (between 19.43\% and 39.48\%), we repeat the imputation procedure 100 times, producing 100 data sets. We apply clustering to each data set and analyze the results to ensure  that the  clustering is robust to the imputation. Specifically, we cluster the subjects in each imputed data set using the popular $k$-means method.

\begin{figure}[ht]
	\centering
	\includegraphics[scale=0.75]{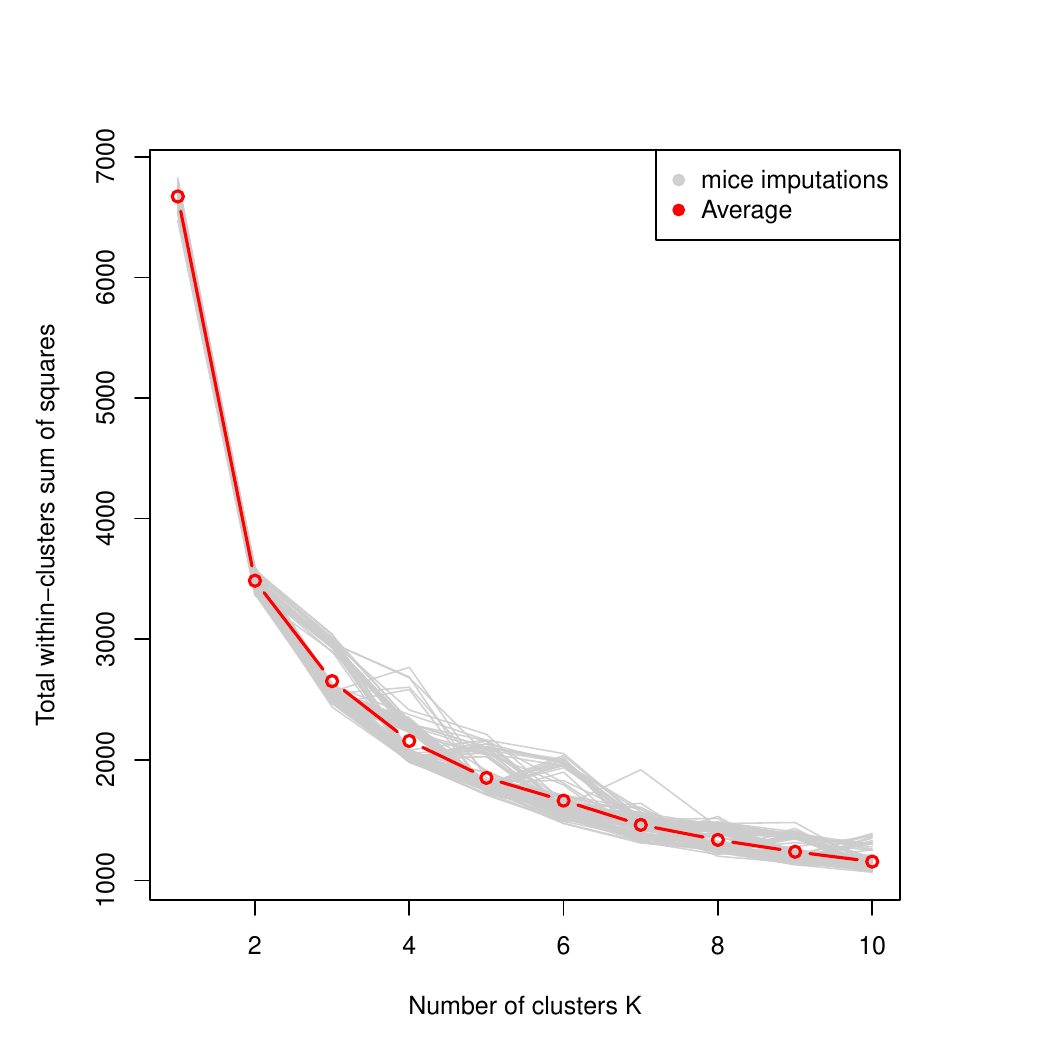}
	\caption{Within sum of squares for different values of the number of clusters $k$ obtained by applying the $k$-means method to 100 imputed $\log_{10}$ LC$_{50}$ data sets. The red curve is obtained as average over the results from the 100 imputed data sets.}
	\label{fig:WSS_kmeans}
\end{figure}

Figure \ref{fig:WSS_kmeans} shows the total within-sum-of-squares (WSS) values obtained for different number of pre-specified clusters (elbow plot). The figure shows the results for each imputed data set in grey, with the average across the 100 imputation data sets in red. It is common practice to select the point where the curve of WSS values in the elbow plot shows the beginning of a plateau as the number of clusters increases. In our case, we set $k = 3$. (We examined larger values of $k$, but the results were similar.) For the main analysis, we select the imputation data set with the lowest WSS value when $k = 3$. In summary, this exploratory analysis is employed to determine the number of clusters for the different drug-sensitivity profiles in the data, as well as to choose an imputed data set to perform the main analysis.  

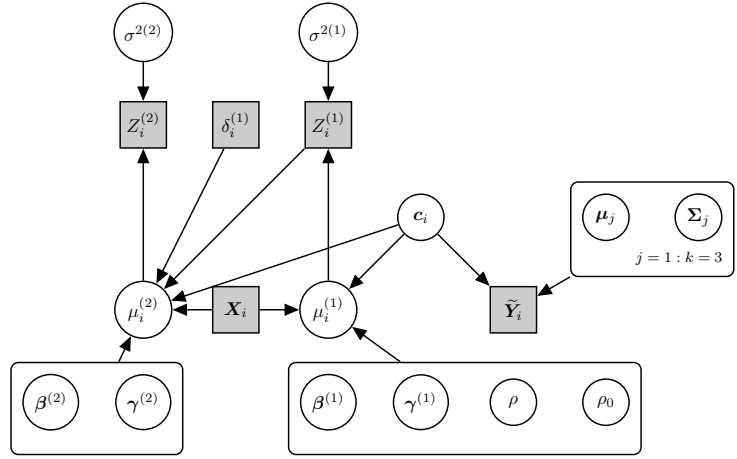
\begin{figure}
	\centering
\subfloat[Model for MRD and LC$_{50}$ observations]{
\tikzset{
	latentnode/.style  ={draw,minimum width=2.5em, shape=circle,thick, black,fill=white},
	visiblenode/.style ={draw, minimum width=2.5em, minimum height=2.5em, shape=rectangle,thick, black,fill=black!20},
	visiblenode2/.style ={draw,minimum width=2.5em, minimum height=2.5em, shape=rectangle,thick, black,fill=white}
}

\scalebox{.7}{
\begin{tikzpicture}[auto,thick,node distance=5em]
	\node[visiblenode] (Z2) {$Z^{(2)}_i$};
	\node[visiblenode, right of = Z2] (deltai) {$\delta^{(1)}_i$};
    \node[visiblenode, right of = deltai] (Z1) {$Z^{(1)}_i$};

	\node[latentnode, above of = Z2] (sig22) {$\sigma^{2(2)}$};
	\node[latentnode, above of = Z1] (sig21) {$\sigma^{2(1)}$};
	
	\node[latentnode, right of = Z1, below of = Z1] (ci) {$\bm c_i$};
	\node[latentnode, below of = ci, left of = ci] (mu1) {$\mu^{(1)}_i$};
	\node[visiblenode, left of = mu1] (Xi) {$\bm X_i$};
	\node[latentnode, left of = Xi] (mu2) {$\mu^{(2)}_i$};

	\node[latentnode, below of = mu2] (gamma2) {$\bm \gamma^{(2)}$};
    \node[latentnode, left of = gamma2] (beta2) {$\bm \beta^{(2)}$};
    \plate [inner sep=0.2cm] {plate2} {(beta2) (gamma2)} {};
	\node[latentnode, below of = mu1] (beta1) {$\bm \beta^{(1)}$};
	\node[latentnode, right of = beta1] (gamma1) {$\bm \gamma^{(1)}$};
	\node[latentnode, right of = gamma1] (rho) {$\rho$};
	\node[latentnode, right of = rho] (rho0) {$\rho_0$};
	\plate [inner sep=0.2cm] {plate1} {(beta1) (gamma1) (rho) (rho0)} {};

	\node[visiblenode, right of = gamma1, above of = gamma1] (Yi) {$\widetilde{\bm Y}_i$};
	\node[latentnode, above of = Yi, right of = Yi] (muj) {$\bm \mu_j$};
	\node[latentnode, right of = muj] (Sigmaj) {$\bm \Sigma_j$};
	\plate [inner sep=0.2cm] {plateYi} {(muj) (Sigmaj)} {$j=1:k=3$};

	\draw [->] (sig22) -- (Z2);
	\draw [->] (sig21) -- (Z1);
	\draw [->] (mu2) -- (Z2);
	\draw [->] (mu1) -- (Z1);
	\draw [->] (plate2) -- (mu2);
	\draw [->] (plate1) -- (mu1);
	\draw [->] (Xi) -- (mu2);
	\draw [->] (Xi) -- (mu1);
	\draw [->] (ci) -- (mu2);
	\draw [->] (ci) -- (mu1);
	\draw [->] (deltai) -- (mu2);
	\draw [->] (Z1) -- (mu2);
	\draw [->] (ci) -- (Yi);
	\draw [->] (plateYi) -- (Yi);
	
\end{tikzpicture}
}
}
\caption{Summary of the proposed models for censored MRD observations and LC$_{50}$ values, highlighting the relationship between the observations and the parameters of the model. White circles represent random variables, while grey squares correspond to observations.}
\label{fig:DAG_Model}
\end{figure}

The selected imputed data set $\widetilde{\bm Y}$ is then modelled via a finite mixture with $k=3$ components as:
\begin{align}\label{eq:Model_Y}
	&\widetilde{\bm Y}_i \mid \left\{ \bm \mu_j, \bm \Sigma_j; j = 1, 2, 3 \right\}, \bm w \ind \sum_{j = 1}^{k = 3}w_j \text{N}_d(\bm \mu_j, \bm \Sigma_j), \quad \bm \Sigma_j = \text{diag}(\sigma^2_1, \sigma^2_2, \sigma^2_3) \nonumber \\
	&\bm w = \left( w_1, w_2, w_3 \right) \sim \text{Dir}(\bm \alpha) \\
	&\bm \mu_j \sim \text{N}_d(\bm 0, \mathbb{I}_d), \quad \sigma^2_j \sim \text{Inv-Gamma}(3, 2), \quad j = 1, 2, 3 \nonumber
\end{align}
where $\text{N}_d(\bm \mu, \bm \Sigma)$ is the $d$-dimensional multivariate  Normal distribution with mean vector $\bm \mu$ and covariance matrix $\bm \Sigma$, while $\text{Dir}(\bm \alpha)$ is the Dirichlet distribution with parameter vector $\bm \alpha$.  We set $\bm \alpha = \left( 1/k, 1/k, 1/k \right)$, corresponding to uniform prior probabilities.
By fitting a mixture model to the imputed dataset, we also obtain an estimate of cluster membership for each individual. That is, using the latent variable representation of a mixture model, we can estimate $\bm c_i\in\{1,2,3\}$, which denotes to which mixture component the $i$-th observation belongs. Clusters correspond to distinct data-driven drug sensitivity profiles in our statistical model. A patient's cluster membership is included in the model~\eqref{eq:MRD_model} through the regression terms $\bm \gamma^{(t)} \bm c_i$, for $t = 1, 2$ and $i = 1, \dots, N$, with cluster membership appropriately coded via a vector of dummy variables $\bm c_i$ of dimension $k-1$. Since $k = 3$, we set the reference category to be $\bm c = (0, 0)$ representing membership to Cluster 1, while $\bm c = (1, 0)$ and $\bm c = (0, 1)$ denote membership to Cluster 2 and Cluster 3, respectively. 
We stress the fact that models~\eqref{eq:MRD_model} and \eqref{eq:Model_Y} are estimated simultaneously, with $\bm c_i$ random and updated within the MCMC algorithm jointly with the remaining parameters. The full model is summarised   in Figure~\ref{fig:DAG_Model}.

\section{Posterior inference}\label{sec:Application}
Posterior inference under the proposed model is obtained via JAGS \citep{plummer2003jags}, interfaced with RStudio via the R package \texttt{rjags} \citep{plummer2019rjags}. We run the MCMC chain  for 15000 iterations, discarding  the first 5000  as burn-in  and  thinning every two.

\begin{figure}[ht]
	\centering
	\subfloat[$(\rho,\rho_0)$]{\includegraphics[width=0.45\textwidth]{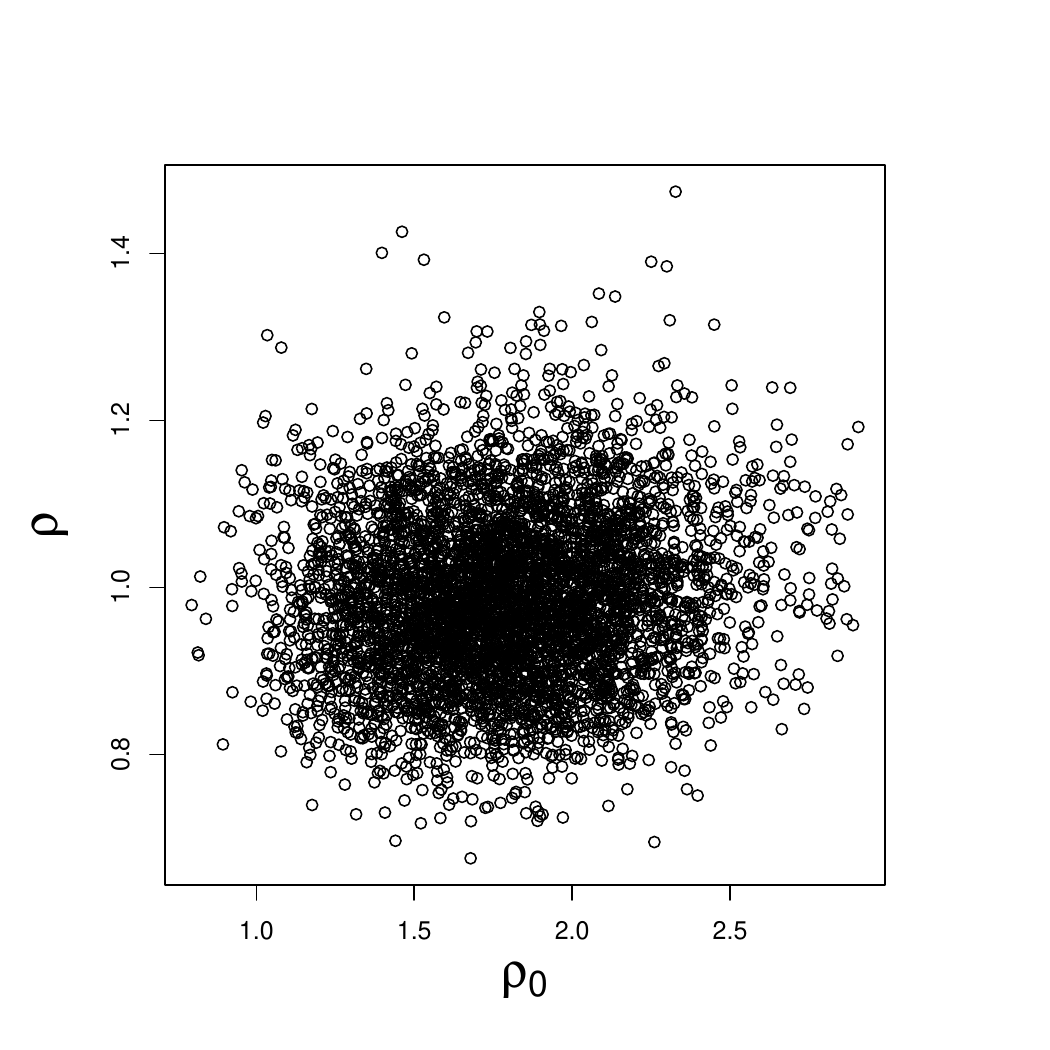}}
	\subfloat[$(\sigma^{2(1)}, \sigma^{2(2)})$]{\includegraphics[width=0.45\textwidth]{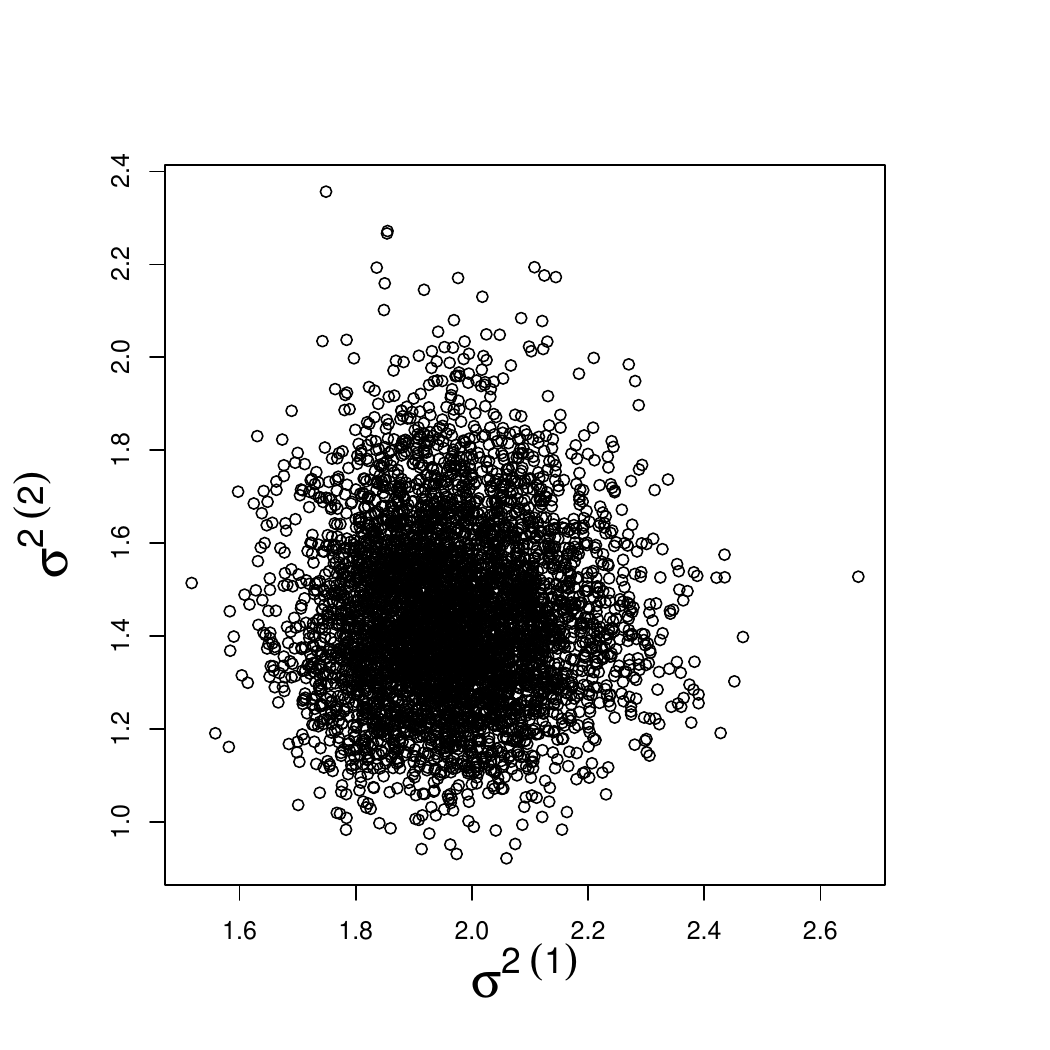}}
	\caption{(a) Posterior distributions of $\rho_0$ and $\rho$; (b) posterior distributions of the variances of the MRD responses at the two time points.}
	\label{fig:Rho_Sigma2}
\end{figure}

Figure~\ref{fig:Rho_Sigma2} displays the scatterplots of the posterior samples of the coefficients $\rho_0$, $\rho$ and of the variances $\sigma^{2(1)}$,  $\sigma^{2(2)}$. 
The posterior distribution of the auto-regressive coefficient $\rho$ is concentrated around the value 1 (posterior median 0.98, posterior interquartile range 0.92-1.05), indicating that the responses at the two different times are strongly associated for those subjects with $\text{MRD}^{(1)} > 0.01$. Furthermore, the residual variances $\sigma^{2(1)}$ and $\sigma^{2(2)}$ show little  \textit{a posteriori} correlation. 

\begin{figure}[ht]
	\centering
	\includegraphics[width=0.75\textwidth]{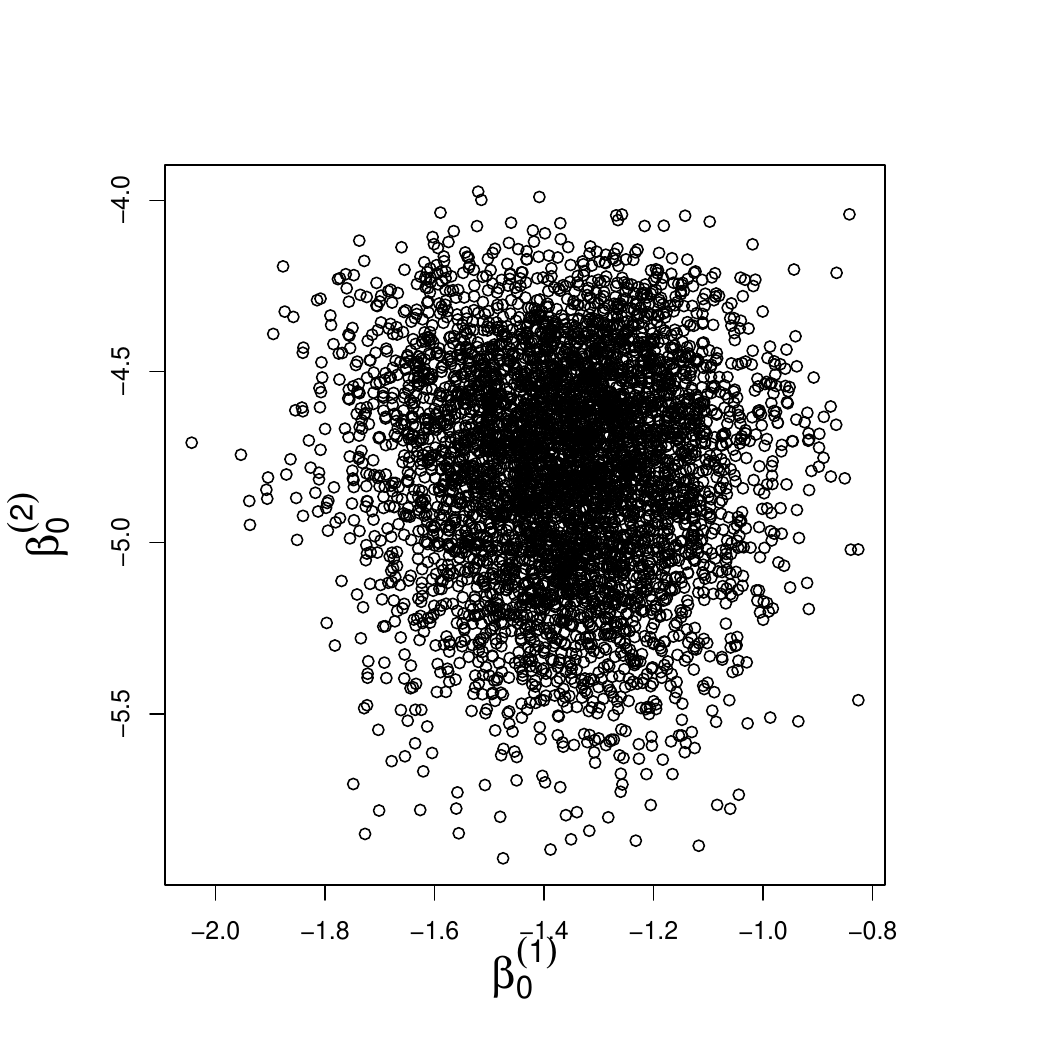}
	\caption{Posterior distributions of the intercept coefficients $\beta^{(t)}_0$, $t = 1, 2$.}
	\label{fig:beta0}
\end{figure}

\begin{figure}[ht]
	\centering
	\subfloat[day 15]{\includegraphics[width=0.5\textwidth]{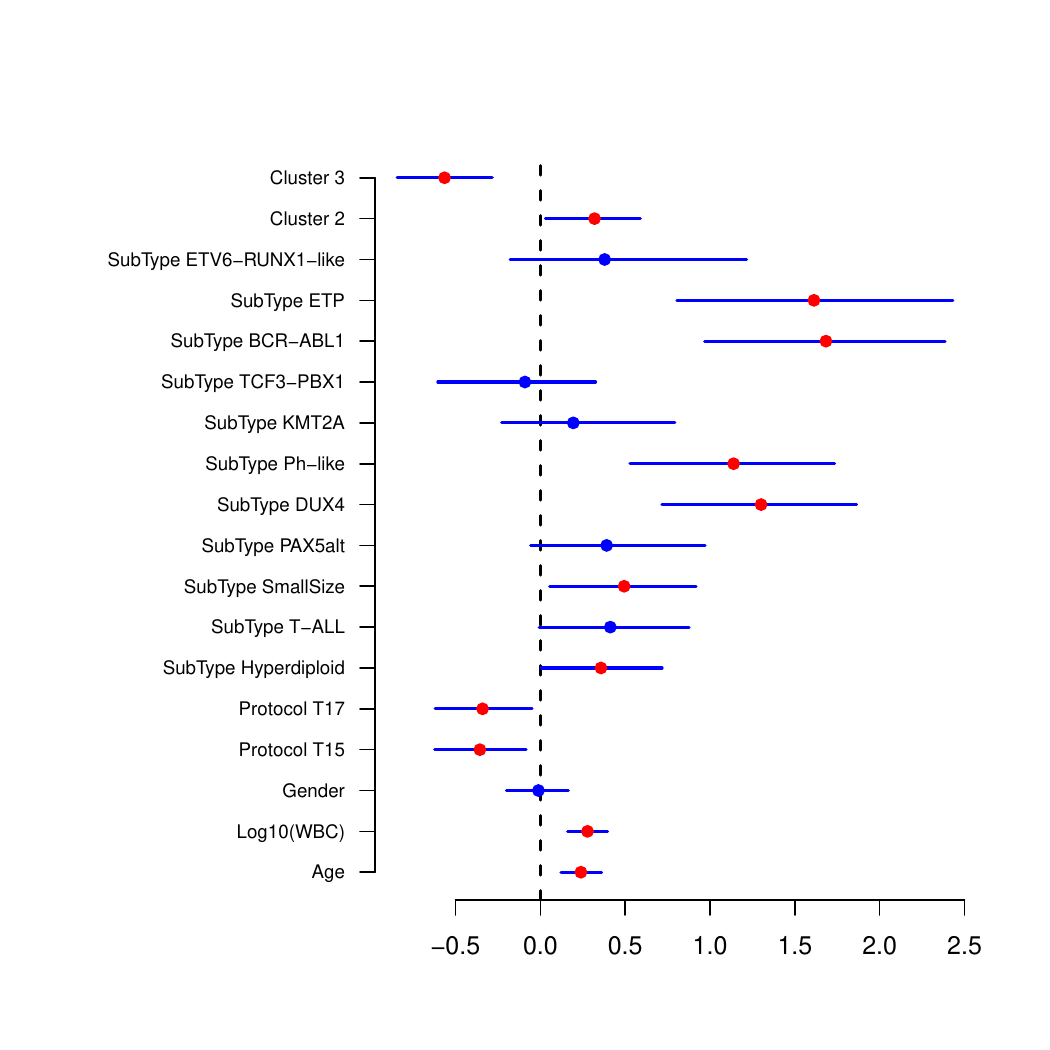}}
	\subfloat[day 42]{\includegraphics[width=0.5\textwidth]{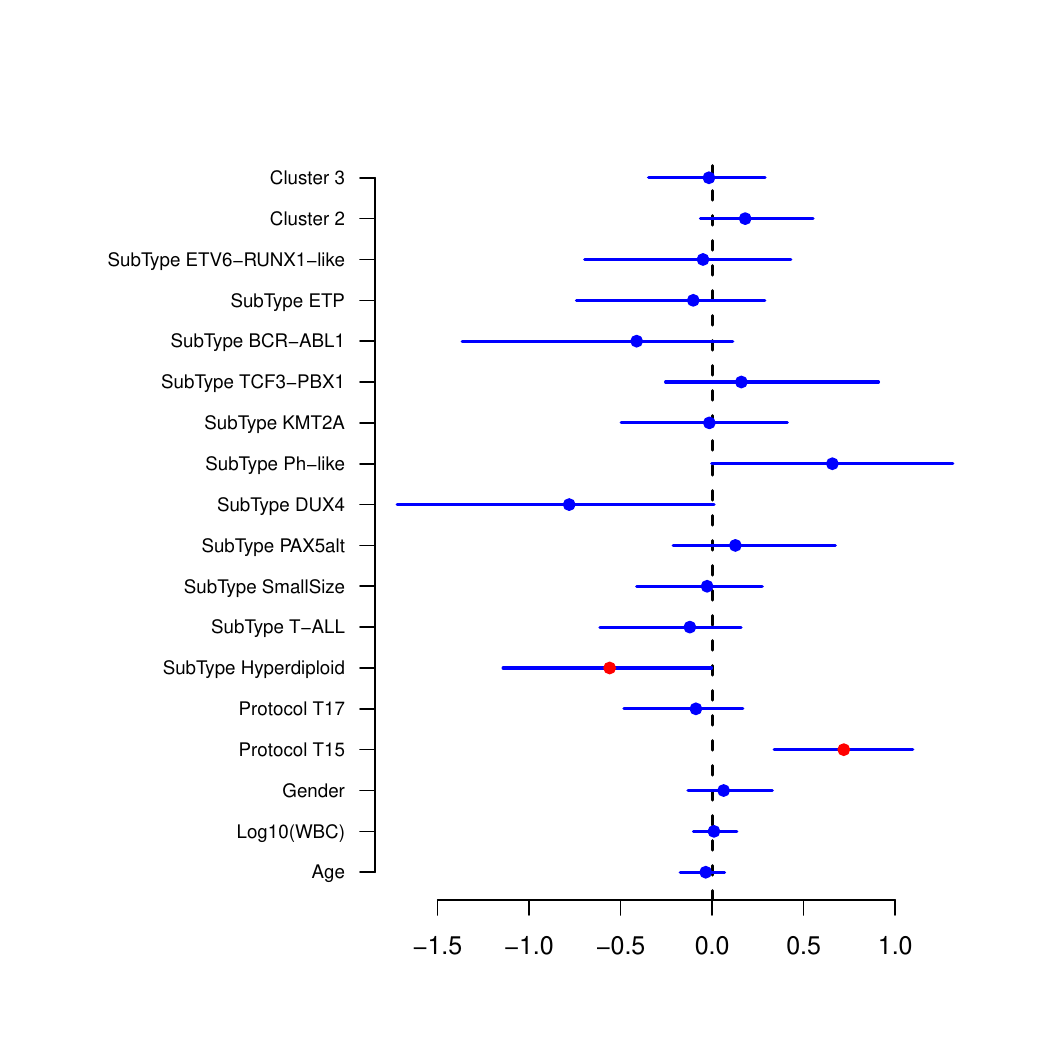}}
	\caption{Posterior distributions of the regression coefficients $\left(\bm \beta^{(t)}, \bm \gamma^{(t)}\right)$ for $t = 1, 2$. The dots represent posterior means of the coefficients. Red colour indicates those regression coefficients whose 95\% credible interval does not include zero, while blue corresponds to those whose 95\% credible interval includes zero. The effects of the categorical variables are reported in relation to their reference categories. The respective reference categories are \textit{Cluster 1} for the sensitivity profile, \textit{ETV6-RUNX1} for the subtype, \textit{T16} for the protocol and \textit{Male} for the gender.}
	\label{fig:betas_gammas}
\end{figure}

We now discuss the effect of the covariates $\bm X_i$ and $\bm c_i$ on the responses. The posterior distribution of the coefficients $\left(\bm \beta^{(t)}, \bm \gamma^{(t)}\right)$ for $t = 1, 2$ are summarized in Figure \ref{fig:betas_gammas}, where we report their posterior means and 95\% credible intervals. Each bar corresponds to one of the covariates included in the analysis. Most of the covariates are associated only with MRD measured at time $t=1$ (i.e., their 95\% credible interval does not contain the value zero). Interestingly, we observe that age, $\log_{10}$ WBC, and protocol are all relevant factors at day 15, while gender is not. The ALL subtype of the patient is also a relevant factor. The subtypes' positive posterior regression coefficients on day 15 show substantially higher MRD on day 15 for these subtypes, relative to the reference subtype \textit{ETV6-RUNX1}, as expressed by many non-zero coefficients. This holds also for the drug sensitivity profiles (see the posterior distribution of variables \textit{Cluster 2} and \textit{Cluster 3}), relative to \textit{Cluster 1}, the reference cluster. We see that \textit{Cluster 2} is associated with lower MRD on day 15 that \textit{Cluster 1}, while \textit{Cluster 3} patients tend to have higher day 15 MRD values than \textit{Cluster 1} patients.
\textit{Cluster 2} has more ETV6-RUNX1 and hyperdiploid patients, which are associated with sensitivity to prednisone and asparaginase, consistent with previous publications.

\begin{figure}[ht]
	\centering
	\includegraphics[width=0.75\textwidth]{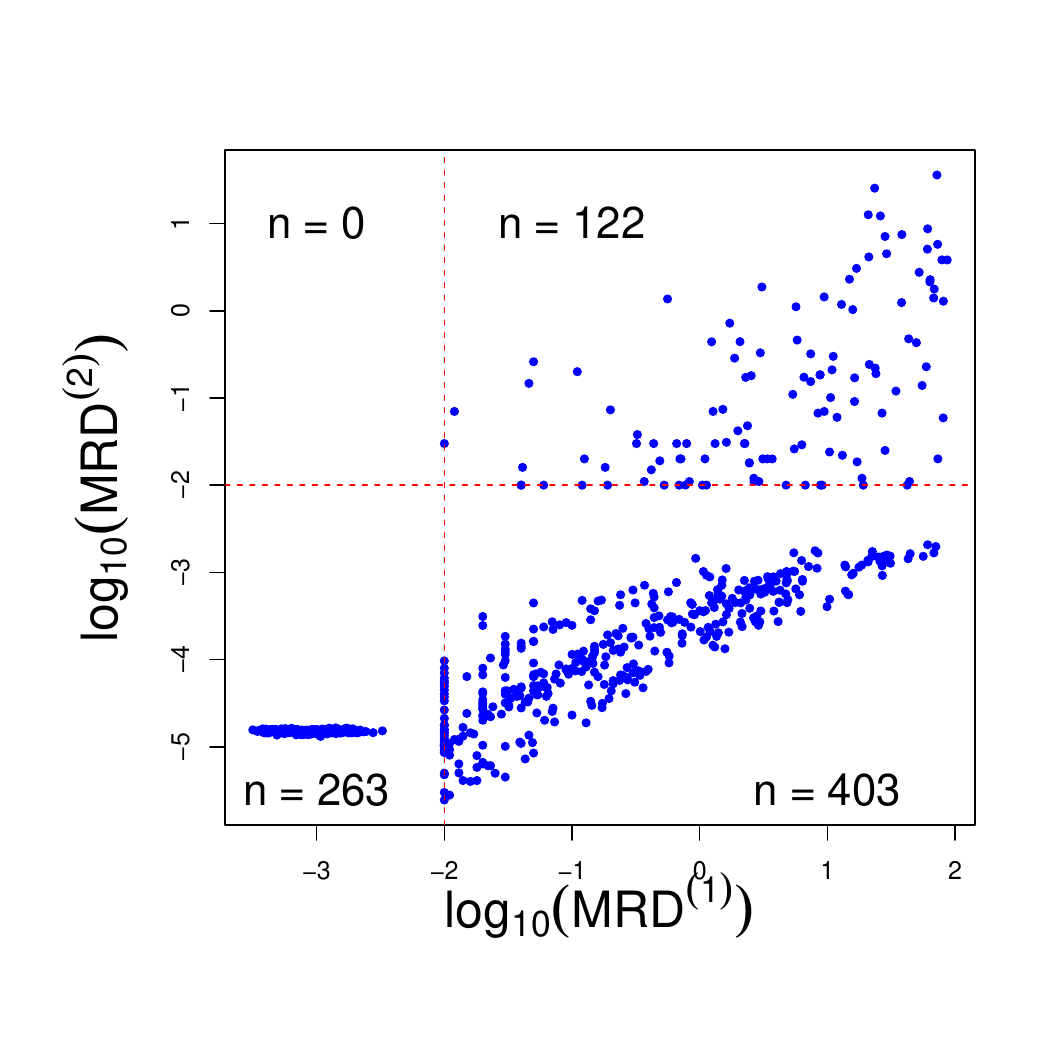}
	\caption{Posterior means of censored and observed MRD values.}
	\label{fig:MRD_est}
\end{figure}

Recall that inference for day 42 depends on the presence of MRD on day 15. We see that Protocol T15 is associated with a higher mean MRD value than the other two studies. T16 and T17 occurred later and included some newer targeted anticancer drugs. We also see an improvement in the effect for the hyperdiploid subtype on day 42, relative to the reference genomic subtype. We also present posterior inference on the censored MRD values, which are imputed within the MCMC algorithm. In Figure \ref{fig:MRD_est} we show the posterior means for the censored observations, as well as the observed MRD values. Note that the bottom left quadrant corresponds to MRD values censored at both time points, while the bottom right panel to observations censored only at day 42. No observation lies in the upper-left quadrant, indicating that none of the subjects in the study experiences a relapse between days 15 and 42.

\begin{figure}[ht]
	\centering
	\subfloat[]{\includegraphics[width=0.325\textwidth]{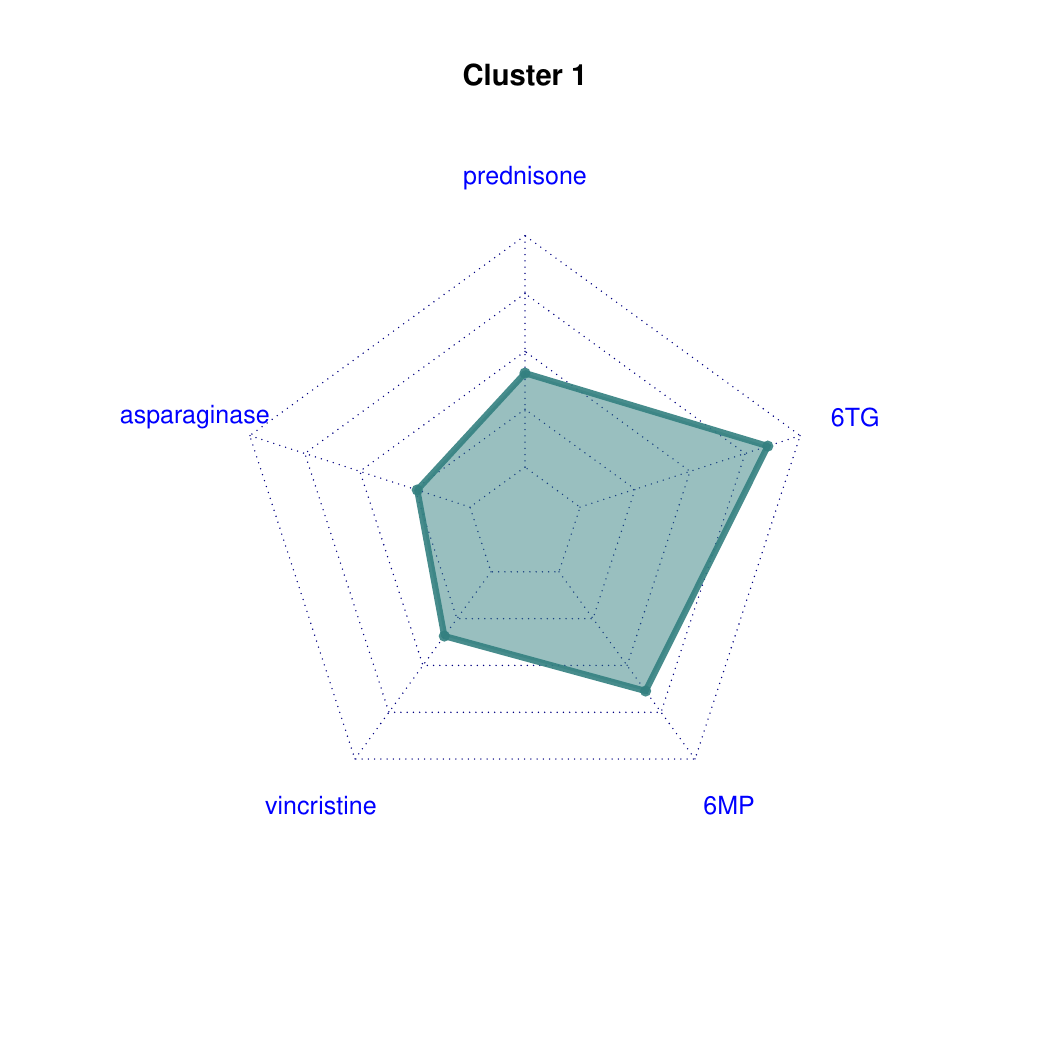}}
	\subfloat[]{\includegraphics[width=0.325\textwidth]{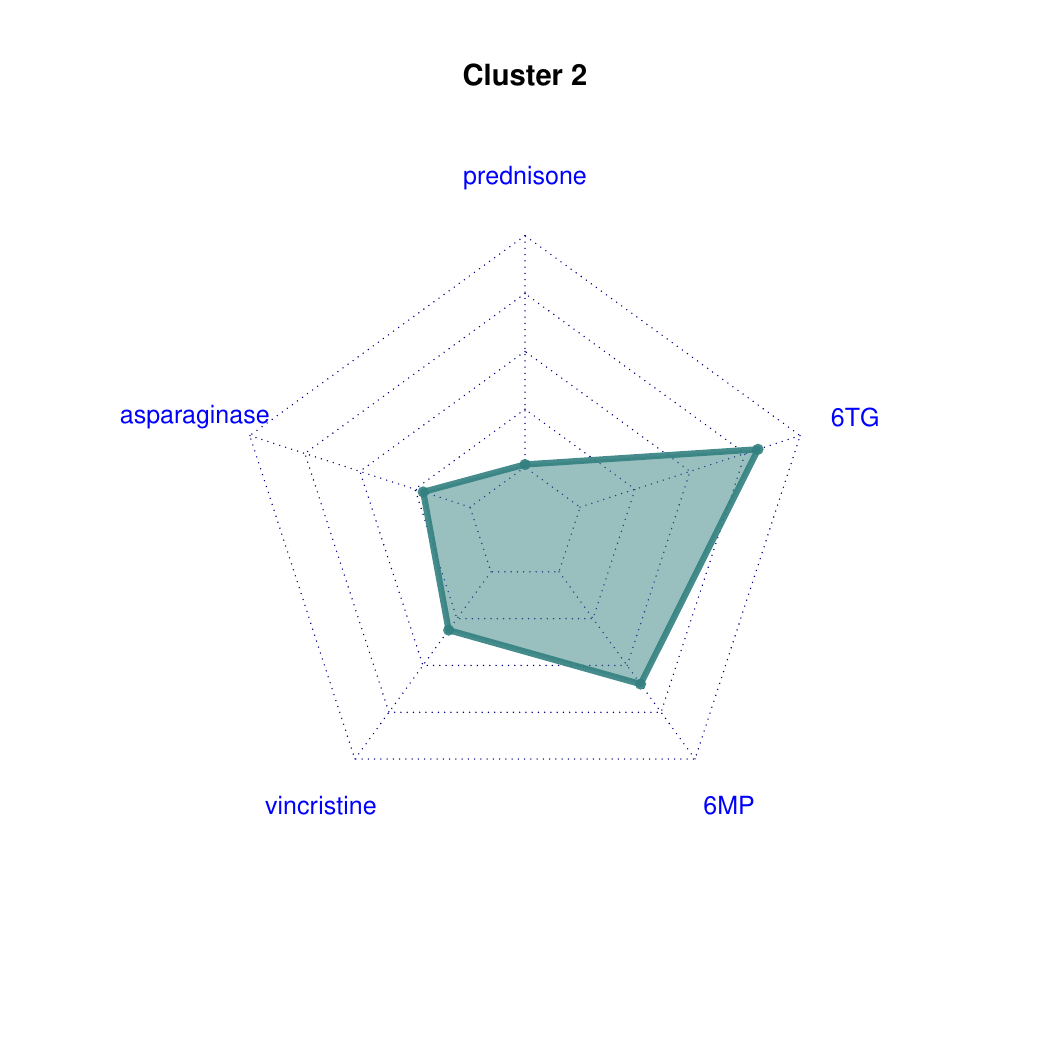}}
	\subfloat[]{\includegraphics[width=0.325\textwidth]{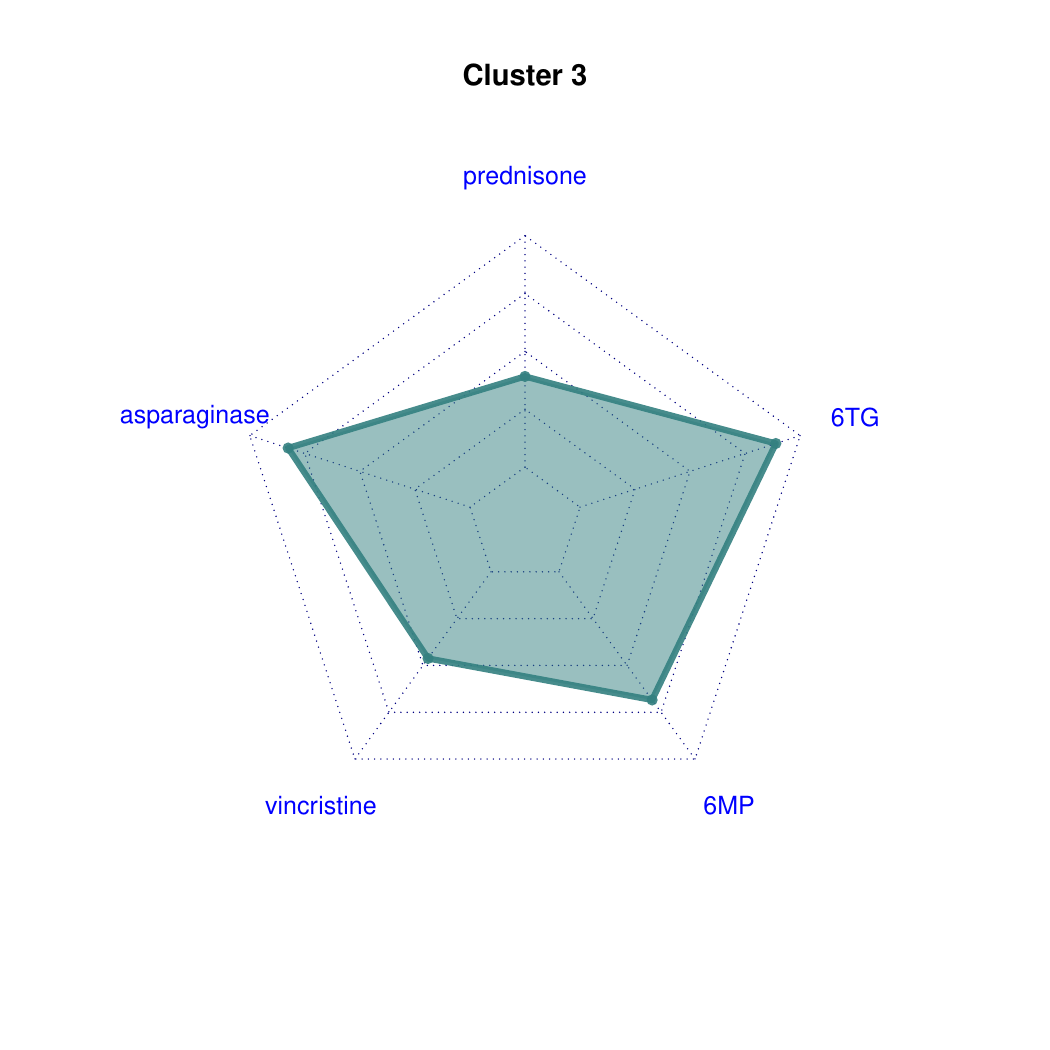}}\\
	\subfloat[]{\includegraphics[width=0.325\textwidth]{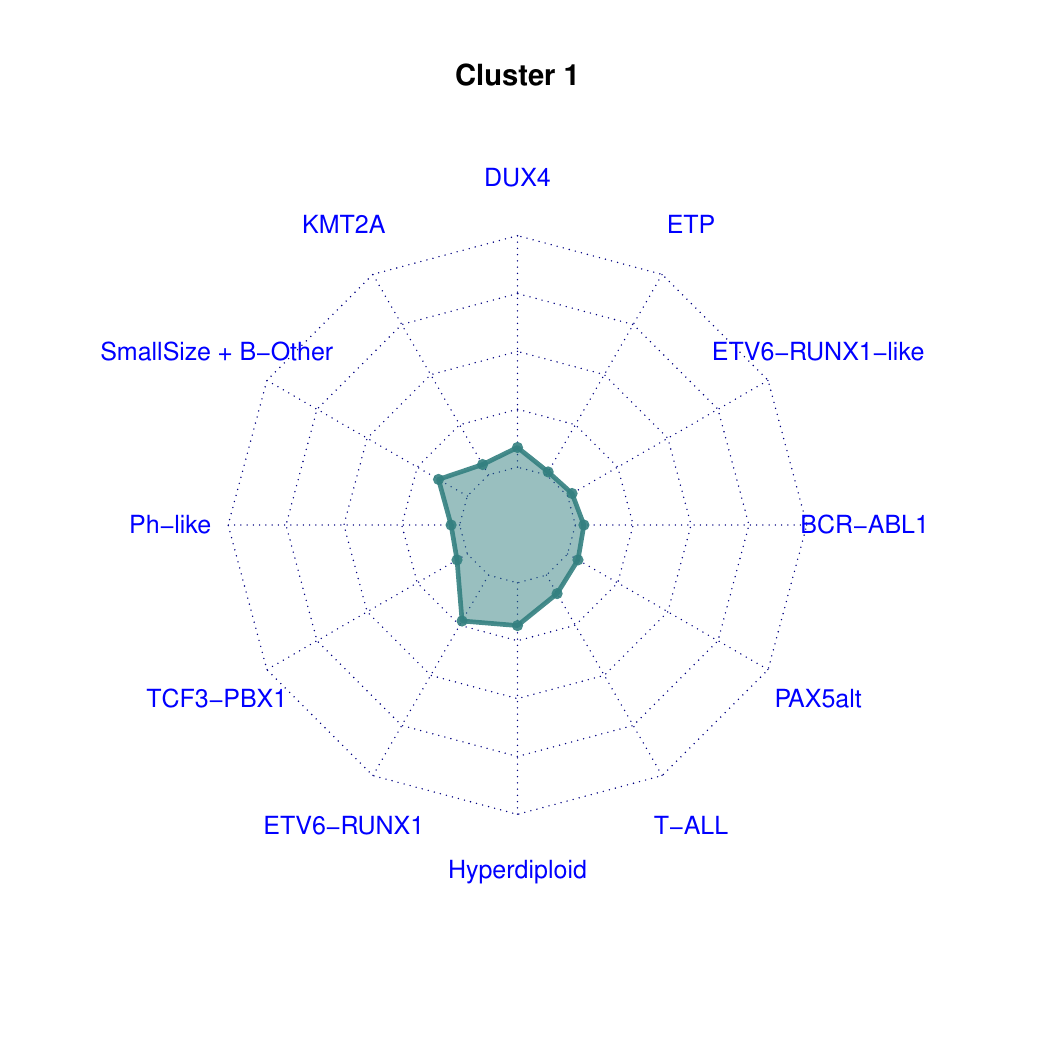}}
	\subfloat[]{\includegraphics[width=0.325\textwidth]{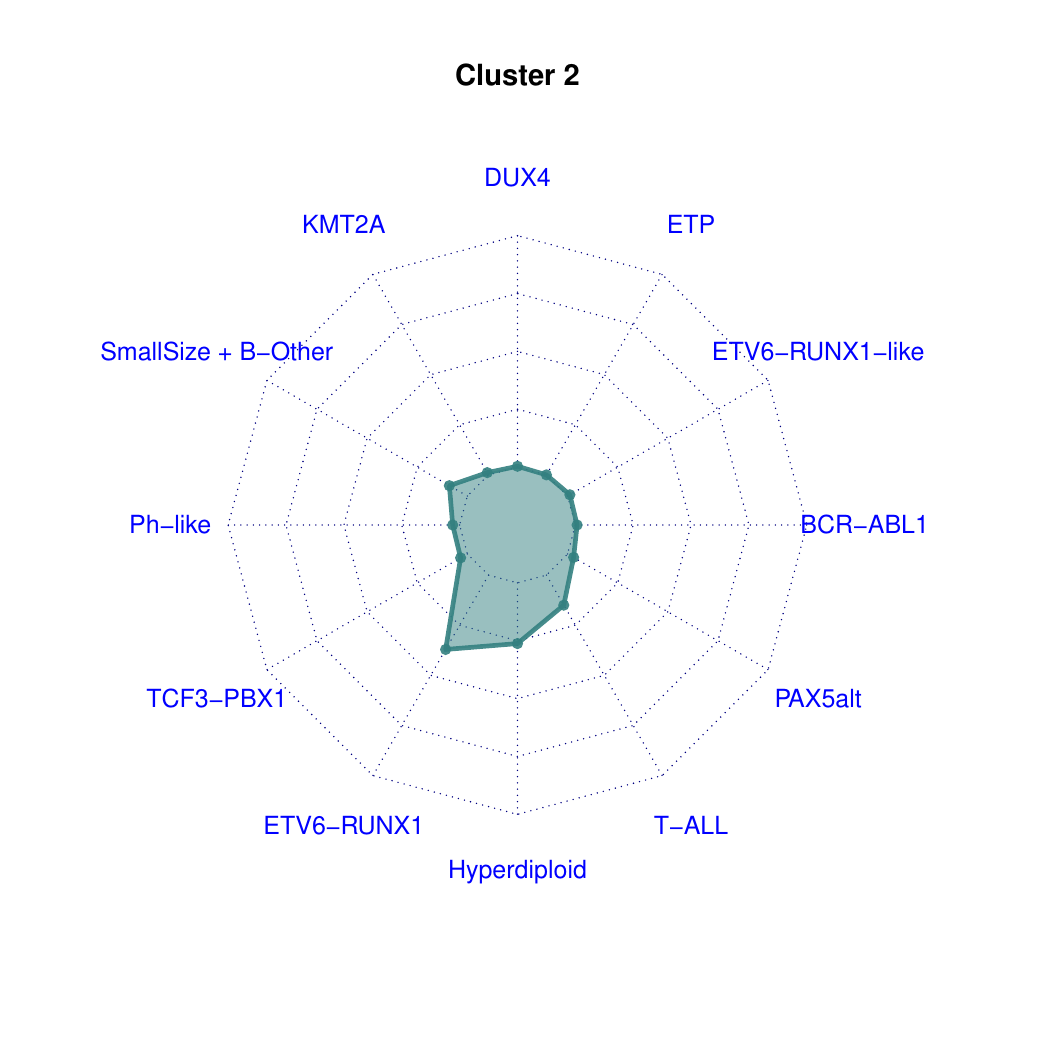}}
	\subfloat[]{\includegraphics[width=0.325\textwidth]{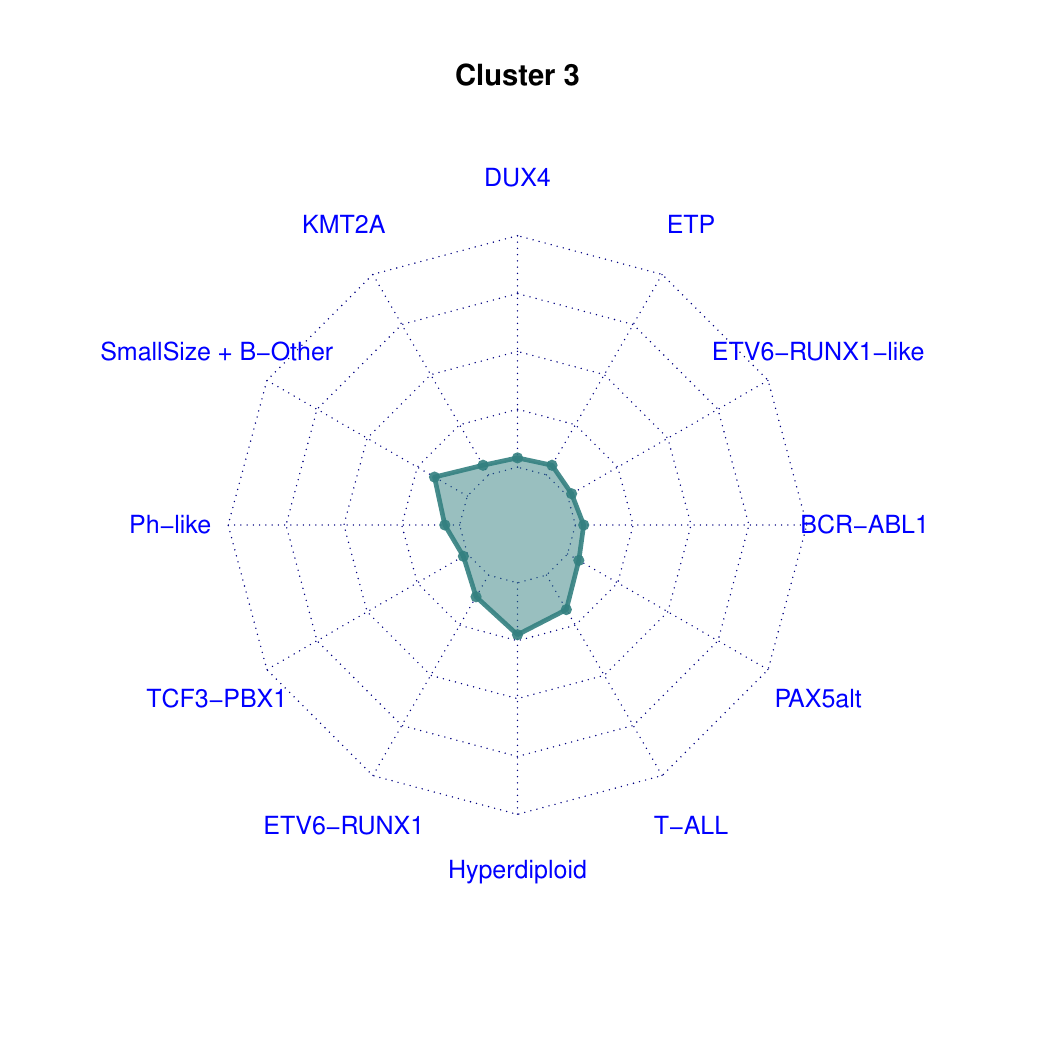}}
	\caption{Radar plots for the average $\log_{10}$ LC$_{50}$ values within each cluster (top row) and for the proportion of subjects within each subtype category (bottom row). Radar plots for $\log_{10}$ LC$_{50}$ values are rescaled such that the vertices of the innermost and outermost polygons (connected by lines) of each cluster's $\log_{10}$ LC$_{50}$ radar plot correspond to minimum and maximum observed values, respectively. The concentric (inner) polygons represent the first, second and third quartiles of the distributions of the observed $\log_{10}$ LC$_{50}$ (top row) or percent of the cluster composition for each subtype (bottom row). The partition of the subjects are estimated by minimizing Binder's loss function.}
	\label{fig:radarplots}
\end{figure}

Clustering of the subjects is achieved via model \eqref{eq:Model_Y}, in which the sensitivity profiles are modelled via a three-component mixture distribution. The posterior estimate of the partition, obtained by minimizing Binder's loss function as implemented in the R package \texttt{salso} \citep{dahl_devin_2020}, presents three clusters of sizes 271, 267 and 250, respectively. We provide an interpretation of the cluster features by investigating the average values of $\log_{10}$ LC$_{50}$s within each cluster, as well as the proportion of subjects characterised by each subtype. The results are summarised in the radar plots in Figure \ref{fig:radarplots}. We observe differences among clusters in the values of the $\log_{10}$ LC$_{50}$ relative to the drugs prednisone and asparaginase, and in the proportion of subjects with subtype ETV6-RUNX1.
\textit{Cluster 2} is associated with the lowest MRD on day 15 (best prognosis). This cluster has more \textit{ETV6-RUNX1} and \JJYDel{H}\JJY{h}yperdiploid patients, which are associated with sensitivity to prednisone and asparaginase, consistent with previous publications (e.g., \cite{HollemanEtAl2004, AutryEtAl2020}). 

\section{Conclusions}\label{sec:Conclusion}

Minimal residual disease  is used in clinical practice to assess  treatment response in patients with ALL. MRD is typically regarded as an independent prognostic factor and is used in clinical trials for risk assignment and to guide clinical therapeutic choices in the management of the treatment of the cancer. In this work we combine information from MRD measurements taken at two time points with patient-specific drug sensitivity estimates and leukemia genomics, with the goal of better understanding treatment response and the role of potential prognostic or predictive (i.e., relevant for treatment selection) factors. Our results highlight  that clustering of individuals in terms of sensitivity to drugs that are part of the induction phase of therapy is mainly driven by sensitivity to two anticancer agents, prednisone and asparaginase. The analysis also highlights the role of genomic subtypes as important determinants of response to induction therapy as measured by MRD. 

From a methodological point of view, the Bayesian framework allows us to jointly estimate the parameters that relate MRD at different time points during the course of therapy to patient-specific covariates, including simultaneously inferred drug sensitivity profiles. Moreover, it is straightforward to account for censoring. A key component of our modelling strategy is that we model MRD at day 42 conditionally on the event that MRD on day 15 is detected. This allows for extra flexibility and accounts for the fact that in our data set remission at day 15 implies remission at day 42. 

A drawback of our modelling strategy is pre-imputation of missing LC$_{50}$ values, which leads to underestimation of uncertainty. We opted for this strategy because of the substantial missing rate. In theory, it is straightforward to impute missing observations within the MCMC algorithm. Future work includes a further investigation of the relationship between subtypes and drug sensitivity. In the data, estimation of LC$_{50}$ occasionally led to estimates below and above the range of concentrations for the patient. Our model did not account for the censoring in estimation of LC$_{50}$s for affected drugs with these patients' samples. 

Another limitation is that the analysis was confined to a subset of ALL subtypes, pooling subtypes with fewer than ten patients. We did consider hierarchical modeling but still pooling. Future work will consider all ALL subtypes as separate categories, including those with small numbers. A robust multi-level model may be useful to account for the uncertainty of the rarer subtypes. Such an extension would be useful, as we can imagine that more and more ``subtypes'' with very few patients will be defined in the future.

Future research will consider the incorporation of a greater number of serial MRD measures in the model. Such a strategy could include measures after remission induction that could be part of monitoring the disease in peripheral blood rather than bone marrow and help fine tune treatment decision rules. 

\clearpage
\bibliographystyle{plainnat}
\bibliography{Biblio}

\end{document}